\documentclass[aps,prc,twocolumn,showpacs]{revtex4-1}
\usepackage{graphicx}
\usepackage{tikz}
\usepackage{subfigure}
\usepackage{amsmath}
\usepackage{epstopdf}
\usepackage{cleveref}
\begin{document}

\title{Hadronization from color interactions}
\author{\small Guang-Lei Li and Chun-Bin Yang}
\affiliation{Institute of Particle Physics \&  Key Laboratory of Quark and Lepton Physics (MOE),
Central China Normal University, Wuhan 430079, People's Republic of China \\
\small E-mail: ligl@mails.ccnu.edu.cn and cbyang@mail.ccnu.edu.cn
}

\date{\today}

\begin{abstract}
A quark coalescence model is presented based on semi-relativistic  molecular dynamics with color interactions among quarks taken into account and applied to $pp$ collisions to study the effects of this model. A phenomenological potential with two tunable parameters is introduced to describe the color interactions between quarks and antiquarks. The interactions drive the process of hadronization and finally make them form different color neutral clusters, which can be identified as hadrons based on some criteria. A Monte Carlo generator, PYTHIA is used to generate the quarks in the initial state of hadronization, and different values of tunable parameters are used to study their effects on the final state distributions and correlations. Baryon to meson ratios, transverse momentum spectra,  pseudorapidity distributions and forward-backward   multiplicity correlations of hadrons produced in the  hadronization process from this model with different parameters are compared with those from PYTHIA.

\end{abstract}


\maketitle

\section{Introduction}

In high energy physics, perturbative Quantum Chromodynamics (pQCD) has been a remarkable success in describing hard processes with large momentum transfers because of the facts that partons behave nearly like free ones and that the interactions can be calculated perturbatively in the process. However, pQCD is invalid  in the long distance regime where the momentum transfers are low.  In this confinement regime, partons with colors are converted into color neutral hadrons through the hadronization process  which is still unable to be solved from first principles due to the non-perturbative feature.
Thus some phenomenological models are often needed to describe the dynamical features of non-perturbative QCD, e.g. the Lund string fragmentation model \cite{mod1,mod2,mod3,mod4,mod5} used in PYTHIA \cite{pyth1,pyth2}, the cluster model used in HERWIG \citep{herwig1,herwig2}, coalescence models \cite{coale1,coale2,coale3,coale4} used in the Monte Carlo transport models such as AMPT \citep{ampt1} and PACIAE \citep{paciae1,paciae2,paciae3}, etc. The Lund model is based on the assumption of linear confinement supported by lattice QCD. In this model, the flux-tube of strong color field between a color charge and its anticolor charge is represented by  a relativistic string which produces a linear confinement potential. The fragmentation process is described by the dynamics of relativistic strings, while the production of quark-antiquark or diquark-antidiquark pairs is modeled as a tunneling process. The cluster model is based on the assumption of pre-confinement. Gluons are splitted into quark-antiquark or diquark-antidiquark pairs after parton shower, followed by the formation of color-singlet clusters which may decay into hadrons isotropically. Coalescence models   connect quark numbers and pre-hadron numbers by assuming an underlying coalescence with some phase space constraint.

The molecular dynamics (MD) method \citep{dy1,dy2} is a microscopic many body approach, which has been successful in materials science, chemical physics and biomolecules \cite{dy3,dy4,dy5}. It has advantages of being able to solve various problems without many assumptions. To study the microscopic dynamics of quark matter during the hadronization process dominated by non-perturbative QCD, the methods of MD  simulation have been applied in recent years \citep{mol1,mol2,mol3,mol4,mol5,mol6,mol7,pol1}.  Ref.  \citep{mol1} studied the the thermalization process of a strongly interacting quark matter. Ref. \citep{mol2} used quantum molecular dynamics to study quark phase transition of a quark system at finite density. A relativistic molecular dynamics approach based on Nambu-Jona-Lasinio (NJL) Lagrangian \cite{njl1} was applied in Ref. \citep{mol3} to study the evolution of high energy heavy ion collisions. The NJL model, as a low energy approximation to the QCD theory, has  already applied to study the properties of baryon-rich matter \cite{njl2,njl3,njl4}. A quark molecular dynamics based on a non-relativistic approach was used to describe the hadronization of an expanding quark gluon plasma in heavy ion collisions with particle creation taken into consideration in Refs. \cite{mol4,mol5}. Refs. \cite{mol6,mol7} analyzed the properties of the quark matter at finite baryon densities and zero temperature by using a semi-classical constrained molecular dynamics approach.

In this paper, we study the hadronization process microscopically using a quark coalescence model in the framework of semi-relativistic molecular dynamics method. In this model, quarks are treated as particles moving synchronously with relativistic velocities  and interacting through a potential, which will lead to the hadronization of quarks. The implementation of the model is based on the work in \cite{pol1} and a generalized form of the Cornell potential used in that paper. In Refs. \cite{mol4,mol5}, the Cornell potential was also applied by using the classical molecular dynamics to yield colorless clusters. However, the details of the implementation, such as the  determination of color factor, the initial state, the algorithm, the criteria to end the evolution and the hadron identification etc., are different from ours. Our study is a preliminary one for now, and mainly focuses on the effects of different parameters on the final state properties of hadrons by analyzing the variations of  baryon to meson ratios, transverse momentum spectra, pseudorapidity distributions and forward-backward multiplicity correlations with different values of parameters.

This paper is organized as follows. In Sec. II, the phenomenological potential and related dynamics are discussed. Its application to the hadronization is described  in Sec. III for quarks produced by PYTHIA for $pp$ collisions. The obtained final state distributions and correlations are presented in Sec. IV. The last one is for some discussions.

\section{The molecular dynamics model for a quark system}

\subsection{The interacting potential}

In our study a Cornell-like phenomenological potential \cite{cornell1,cornell2,pot1} is employed with some free parameters. The Cornell potential has gained success in describing the properties of bounded states, especially heavy quark mesons \citep{pot1,pot2,pot3,pot4}. It is parametrized as a linear combination of the Coulombic and linear potentials. The Coulombic term, similar to the electromagnetic case in behavior, is responsible for the one-gluon exchange between two quarks, while the linear term dominates  when the distance is large and thus gives rise to confinement. In this paper, the quarks and antiquarks are treated as  particles interacting through this kind of potential depending on their distance and color combination. The potential between a pair of quarks $i$ and $j$ separated by a distance $r_{ij}$ is defined as
\begin{eqnarray}
\label{equ1-1}
V_{ij}=\alpha_{ij}(ar_{ij}-\frac{b}{r_{ij}})+c \ ,
\end{eqnarray}
where $a$, $b$ are coefficients that describe the strength of interactions, $c$ is a constant, and $\alpha_{ij}=\alpha_{ji}$ is a color factor determined by the combination of color charges (red, green or blue for quarks, and their anticolors for antiquarks) of these two quarks. The units for $a$, $b$, and $c$ are GeV/fm, GeV $\cdot$ fm, and GeV respectively.

This paper  regards $a$ and $b$ as adjustable parameters. $a$ and $b$ correspond to the string tension and phenomenological strong coupling constant respectively. Note here the unit of  $b$  is GeV $\cdot$ fm, which needs to be converted to get the usual dimensionless coupling constant, e.g. $b=0.1$ GeV $\cdot$ fm $\approx 0.508$. However, considering our main concern is the influence of the change of $a$ and $b$ on the hadronization process, their absolute values are not very relevant. We set $a>0$. For given $a$ and $b$, $c$ can be determined by supposing the total potential energy between all pairs of quarks and antiquarks in the initial state is zero to conserve the total  energy.

The value of the color factor $\alpha_{ij}$ for a combination of color charges includes two parts: the sign for the combination of color charges and the relative strength of the color factor.

The sign of $\alpha_{ij}$ for a color combination can be determined by considering different kinds of combinations of quarks carrying color charges. For a quark and an antiquark, if they carry one color charge and its anticolor charge respectively, they can form a color neutral meson, then they should attract each other, and thus $\alpha_{ij}>0$; otherwise, the force interacting between them is repulsive, and $\alpha_{ij}<0$. For two quarks (or antiquarks), if their colors are different, they  may form a diquark (or an antidiquark), then their interaction is attractive and $\alpha_{ij}>0$; otherwise their interaction is repulsive and $\alpha_{ij}<0$.

\begin{table}[!htbp]
\centering
\begin{tabular}{|c|c|c|c|c|c|c|}
\hline\hline
$\alpha_{ij}$  & $r$ & $g$ & $b$ & $\overline{r}$  & $\overline{g}$  & $\overline{b}$ \\
\hline
$r$ & -1 & $\frac{1}{2}$  & $\frac{1}{2}$ & 1 & -$\frac{1}{2}$ & -$\frac{1}{2}$ \\
\hline
$g$ & $\frac{1}{2}$ & -1 & $\frac{1}{2}$ & -$\frac{1}{2}$ & 1 & -$\frac{1}{2}$ \\
\hline
$b$ & $\frac{1}{2}$ & $\frac{1}{2}$ & -1 & -$\frac{1}{2}$ & -$\frac{1}{2}$ & 1 \\
\hline
$\overline{r}$  & 1 & -$\frac{1}{2}$ & -$\frac{1}{2}$ &  -1 &  $\frac{1}{2}$ & $\frac{1}{2}$ \\
\hline
$\overline{g}$  & -$\frac{1}{2}$ & 1 & -$\frac{1}{2}$ & $\frac{1}{2}$ &  -1 & $\frac{1}{2}$\\
\hline
$\overline{b}$  & -$\frac{1}{2}$ & -$\frac{1}{2}$ & 1 & $\frac{1}{2}$ & $\frac{1}{2}$ & -1\\
\hline\hline
\end{tabular}
\caption{The color factor $\alpha_{ij}$ for different combinations of color charges. }
\label{tab1}
\end{table}

Next is the determination of relative strength of $\alpha_{ij}$ for different combinations of color charges. The net force acting from a hadron with neutral color on a quark or antiquark should be zero if the hadron can be regarded as a point particle. Suppose that is a result of the offsetting of forces from the quarks composing the hadron, then the relative strength of different combinations of color charges can be determined by comparing the forces acting from  the quarks in a meson or baryon on a quark (or antiquark) outside. For a quark and an antiquark comprising a meson, their colors must be complemented. If the color charge a quark outside carries is the same as that of a quark inside the hadron, then the force acting from the same color quark on the outside one is repulsive, and the force acting from the antiquark inside the meson is attractive. Then one can conclude that the absolute value of relative strength of $\alpha_{ij}$ between the combination of same color charges and the one with complemented color charges is 1:1. If the color charge of an outside quark is different and non-complementary color to that of a quark inside a meson, the absolute value of relative strength of attraction between different color charges and the repulsion between non-complemented color charges is also 1:1.   (The same results can be obtained by supposing an antiquark outside). In the case of a quark outside a baryon composed of three quarks, the color of the outside quark must be the same as that of one quark in the baryon, and different from the other two, then the relative strength of repulsive force between quarks of same color should be twice  the attractive force between quarks with different colors. Other combinations can be considered likewise. In this way, the relative strength between different combinations of color (or anticolor) charges can be deduced. Finally, combining with the signs of $\alpha_{ij}$, the relative values of $\alpha_{ij}$ can be summarized in Table \ref{tab1}.

The color factors deduced here are the same as these in Refs. \cite{mol4,mol5}, which are obtained from the Lagrangian for one gluon exchange of the QCD interactions.

\subsection{Dynamics}

Since it is  hard to describe the microscopic dynamics of a relativistic many body system in a relativistically consistent way \citep{rel1,rel2}, in this paper we treat quarks and antiquarks as particles whose motions are driven by the forces they exert on each other. For a system consisting of quarks and antiquarks, the color charges of quarks (or antiquarks) are fixed as an approximation. Additionally, gluons are not considered as particles in the model, but are accounted for the background field in the system.

It should also be noted that the relativistic effects, such as retardation and chromomagnetism should be taken into account, since most quarks in our simulation are light quarks. However, because of many difficulties for many-particle system, these are not included in our present work. Additionally, since the the proper time of each particle in the system is different from others, the choice of  reference frame is also an issue. In our study, we only consider the time with respect to the laboratory frame   (i.e. the center of mass frame of the $pp$ collision), and assume all quark and antiquarks interact synchronously.

The motions of quarks and antiquarks in the system are influenced by their mutual interactions. From Eq. (\ref{equ1-1}), one can get the force acting on quark $i$ from quark $j$,

\begin{eqnarray}
\label{equ2-1}
\vec{F_{ij}}=-\nabla V_{ij}=-\alpha_{ij}(a\frac{\vec{r}_{ij}}{|\vec{r}_{ij}|}+b\frac{\vec{r}_{ij}}{|\vec{r}_{ij}|^3})\ ,
\end{eqnarray}
and the total force acting on quark $i$ from the rest quarks and antiquarks in one system is
\begin{eqnarray}
\label{equ2-2}
\vec{F_{i}}= \sum_{j\neq i}-\alpha_{ij}(a\frac{\vec{r}_{ij}}{|\vec{r}_{ij}|}+b\frac{\vec{r}_{ij}}{|\vec{r}_{ij}|^3}) \ .
\end{eqnarray}

The non-covariant equations of motion of the quark $i$ are

\begin{equation}
\label{equ2-3}
\left \{
\begin{aligned}
\frac{d\vec{r}_i}{dt} &=\vec{v}_{i} \ , \\
\frac{d\vec{p}_i}{dt} &=\vec{F}_{i} \ , \\
\vec{v}_i &=\frac{\vec{p}_i }{\sqrt{{m_i}^2+{\vec{p}_i}^2}} \ .
\end{aligned}
\right.
\end{equation}

Eqs. (\ref{equ1-1}) and (\ref{equ2-1}) have no meaning if $r_{ij} \to 0$. To avoid this, Ref. \cite{mol1} introduced a short-distance cutoff, i.e. at $r_{ij} < 0.1 \rm \ fm$, the potential was taken as a linear function of $r$, which was explained as a result of the finite spatial extension of the quark wavefunction. Our model sets a minimum distance $r_{\rm min}=0.1 \rm \ fm$ between quarks. A ``contact'' interaction would happen if the distance between quarks $i$ and $j$ is less than $r_{\rm min}$ and they are approaching each other. The interaction between them is implemented by assuming that the force is along the line connecting the two quarks (If there are more than two quarks with distance less than $r_{\rm min}$, one can let the two quarks  with minimum distance have a ``contact'' interaction and bounce back first, and consider the other pairs consequently). As a result, the components of momenta perpendicular to $\vec{r}_{ij}$ are unchanged after the collision, whereas the components of momenta parallel to $\vec{r}_{ij}$ can be solved by considering the conservation of energy and momentum \cite{rel2,rel3},

\begin{eqnarray}
\label{equ2-6}
\vec{p}_{i,\parallel}^{'} &=& \gamma (\vec{v}_{cm})^2[2\vec{v}_{cm}E_i-(1+|\vec{v}_{cm}|^2)\vec{p}_{i,\parallel}] \ , \nonumber \\
E_{i}^{'} &=& \gamma (\vec{v}_{cm})^2[(1+|\vec{v}_{cm}|^2)E_i-2\vec{v}_{cm}\cdot \vec{p}_{i,\parallel}] \ ,
\end{eqnarray}
where the variables with superscript $'$ denote the corresponding ones after collision, and $\vec{v}_{cm}=(\vec{p}_{i,\parallel}+\vec{p}_{j,\parallel})/(E_i+E_j)$ is the collision invariant velocity of the center of mass of these two quarks.

\subsection{Integration method}

The molecular dynamics simulation is performed through dividing the whole evolution time into enormous number of tiny timesteps. For each step, if the force acting on each quark and the velocity of each quark are regarded as constant ones, then the position and momentum  for next time can be calculated according to Eqs. (\ref{equ2-3}) by using some kind of integration method. In this way the time advances a step, the position and momentum of each quark are updated, and then the above process can be repeated.

Apparently, if the timestep is larger, the error of integration is larger, which leads to the relatively larger error of dynamical properties and energy drift. On the other hand, if the timestep is too small, the cost of simulation will be too large, because for each step, updating the positions and momenta of all quarks is time-consuming. especially the updating of forces. For each step, the updating of forces needs to calculate the relative distances between all pairs of quarks in the system, which is quite time-consuming if the number of quarks is large. Thus, the integration algorithm is vital to the accuracy and efficiency of simulation.

The velocity Verlet algorithm \cite{dy1} is adopted in our model, considering that its accuracy is of order two (which is better than the usual Euler method) and it consumes less memory since each iterative step just needs the properties of the former one. In addition, it only needs to update forces once for each step.

Suppose the position, momentum, kinetic energy (rest energy included) of quark $i$ at time $t$ is $\vec{r}_i(t)$, $\vec{p}_i(t)$, $E_i(t)$, the force acting on it is $\vec{F}_i(t)$, and the timestep is $\delta t$. The corresponding ones at next step $t+\delta t$ can be determined as follows. The momentum of quark $i$ at $t+\frac{1}{2}\delta t$ is

\begin{eqnarray}
\label{equ3-1}
\vec{p}_i(t+\frac{1}{2} \delta t)=\vec{p}_i(t)+\frac{1}{2}\delta t \vec{F}_i(t)\ ,
\end{eqnarray}
from which one can get the velocity at $t+\frac{1}{2}\delta t$,
\begin{eqnarray}
\label{equ3-2}
\vec{v}_i(t+\frac{1}{2}\delta t) =\frac{\vec{p}_i(t+\frac{1}{2}\delta t)}{\sqrt{m_i^2+{\vec{p}_i(t+\frac{1}{2}\delta t)}^2}}\ .
\end{eqnarray}

Then the position at $t+\delta t$ can be written as
\begin{eqnarray}
\label{equ3-3}
\vec{r}_i(t+\delta t) =\vec{r}_i(t)+\vec{v}_i(t+\frac{1}{2}\delta t)\delta t\ .
\end{eqnarray}

After the update of position of each quark, the forces acting on each quark at $t+\delta t$ can be obtained using Eq. (\ref{equ2-2}), which can be used to calculate the momentum at $t+\delta t$

\begin{eqnarray}
\label{equ3-4}
\vec{p}_i(t+\delta t) =\vec{p}_i(t+\frac{1}{2}\delta t)+\frac{1}{2}\delta t \vec{F}_i(t+\delta t)\ .
\end{eqnarray}

In this way, the coordinates and momenta of quarks in a system are advanced one timestep $\delta t$.

\section{Influence of interactions on the hadronization}

One can apply the potential model above to study the process of hadronization for quark systems produced by PYTHIA for $pp$ collision at different energies using the molecular dynamics method. PYTHIA 8.230 is used in the calculation.

\subsection{Initial state}

The initial state of quarks for a $pp$ collision event is generated from a Monte Carlo generator, PYTHIA \cite{pyth1,pyth2}.

PYTHIA is a simulation program that can be used to generate events in high-energy collisions between two incoming particles (e.g. $pp$, $ep$ and $e^{+}e^{-}$). It is based on a series of analytical results and a coherent set of models based on QCD from a few body hard process to a multiparticle final state. Its physics features consist of hard and soft interactions, initial/final-state showers, beam remnants, multiple parton interactions, fragmentation and decay. The PYTHIA 8.2  version \cite{pyth2} used in our model is a complete rewrite from Fortran to C++ and can be used for experimental or phenomenological studies, especially for the LHC studies. $pp$ collision is decomposed into parton-parton collisions in this model, which can be divided into soft and hard collisions, wherein the hard part is described by  the lowest leading order perturbative QCD, whereas the soft part is considered empirically.

Hadrons produced directly by hadronization process of PYTHIA model (i.e. hadrons before decay) are input in our model as ``parent hadrons'' and their identities are marked using the codes according to ``The Monte Carlo particle numbering scheme'' \cite{pdg}  used in PYTHIA. Then they are decomposed into quarks and antiquarks according to the flavors and spins of their velence quarks. A meson is converted to a quark and an antiquark, while a baryon (or antibaryon) is first converted to a quark and a diquark (or an antiquark and an antidiquark), and the latter one again is broken into quarks (or antiquarks). This decomposition is assumed to be isotropic in the rest frame of the parent hadron. Details of the implementation of decomposition are the same as those of the diquark break-up method in Ref. \cite{paciae1}. This approach of obtaining initial state is similar to that in AMPT with string melting \cite{ampt1}. The  decomposing of ``parent hadrons'' is equivalent to keeping the evolution process at the parton level until all quarks and antiquarks are produced in the string fragmentation process in PYTHIA.

The sum of color charges of quarks from the same hadron is kept neutral, then the whole system is also color neutral.

The masses of quarks and antiquarks during this decomposition are taken to be current masses used in PYTHIA model, e.g. $m_d=0.0099 \ \rm GeV/c^2$, $m_u=0.0056 \ \rm GeV/c^2$, $m_s=0.199 \ \rm GeV/c^2$, $m_c=1.23 \ \rm GeV/c^2$, $m_b=4.17 \ \rm GeV/c^2$. The constituent masses are not used in this process because for a $0^{-}$ pseudoscaler meson, as Goldstone boson, its mass may be smaller than the sum of mass of constituent quarks, which makes it impossible to convert these mesons into quarks.

After all hadrons from PYTHIA are decomposed, an initial state consisting of quarks and antiquarks is generated. The properties, like masses, flavors, colors, momenta of all quarks and antiquarks are determined. However the positions are not,  since the  information of coordinates of parent hadrons from PYTHIA is absent.

Here we set the initial positions of quarks and antiquarks using a simple method as follows. Firstly, we assume the parent hadrons are uniformly distributed in a circle lying in a transverse plane with zero longitudinal coordinate. This is similar to the assumption used in Refs. \citep{resc1,resc2} by considering that a $pp$ collision can be approximately regarded as a disc as a result of Lorentz contraction in the laboratory frame. The circle's radius is set to be 1 fm. Then define the formation time $t_{f}$ of one quark as the time traveling from the point where the parent hadron is generated to the position where it is decomposed from its parent hadron. The parent hadron moves along a straight trajectory with a constant velocity. This method sets  initial positions of quarks by introducing a formation time and using straight-line trajectories of their parent hadrons. It is similar to that in the model of AMPT with string melting, but doesn't consider the transverse momentum dependence of the formation time. It also doesn't take into account the  Lorentz boost effect of the formation time, otherwise the quarks with large momenta will be spaced too far apart. If now the interactions of the system begin, then quarks and antiquarks from the same hadron would  immediately return to the original hadron combinations. For this reason, we suppose the quarks decomposed from parent hadrons do not interact until they have moved an additional time $t_f^{'}$ with their respective constant velocities. $t_f^{'}$ is referred to as free flow time of quarks. Then the position of a parton at the beginning of evolution is
\begin{eqnarray}
\label{equ4-4}
\vec{r}_i =& \vec{r}_0+\vec{v} _h t_f+\vec{v} _q t_f^{'}
=& \vec{r}_0+ \frac{\vec{p}_h}{E_h}t_f+\frac{\vec{p}_q}{E_q}t_f^{'}\ ,
\end{eqnarray}
where $\vec{r}_0$ is the initial position (whose longitudinal coordinate is zero) of the parent hadron of the quark.

Considering that the quarks are formed at different stages during one $pp$ collision process, we suppose the formation time is a uniformly distributed random variable between 0 and a maximum formation time $t_{f,\rm max}$. As revealed from the results presented below, the final distributions of hadrons  are strongly influenced by the initial geometry of quarks. Given the absence of sensible method of determination of initial coordinates, the method above is just used as a first attempt.

In fact, according to the calculation from the initial geometry above, the value of $c$ for the parameter in the interaction potential in Eq. (\ref{equ1-1}) is much smaller than $a$ and $b$ (about two orders of magnitude smaller), especially for events with large quark multiplicity.

Two kinds of quark masses are used in our model. In addition to the current masses used during the decomposition of parent hadrons mentioned above, during the evolution process, quark masses are taken to be the constituent masses, $m_d=0.325\rm GeV/c^2$, $m_u=0.325 \ \rm GeV/c^2$, $m_s=0.5 \  \rm GeV/c^2$, $m_c=1.6 \ \rm GeV/c^2$, $m_b=5.0  \ \rm GeV/c^2$. The current masses are changed into constituent masses of their respective quarks after the decomposition. The differences between these two kinds of masses lead to an energy discrepancy because of the change of rest energy of each quark, which is negligible compared with the energy of the collision system.

\subsection{Evolution and hadronization}

Because the speed of a quark with large momentum (e.g. over 100 GeV) is usually very close to that of light, and its momentum is nearly equal to its energy in magnitude. Thus the change of momentum doesn't have much effect on its change of velocity and it's quite hard to separate quarks into color neutral clusters with speeds very near to 1. Therefore, in our study the dynamics of high energy quarks (decomposed from parent hadrons with energy larger than 50 GeV) are excluded from the evolution simulation for now.

After removing high energy quarks,  we use Verlet method mentioned above to simulate the numerical evolution of the quark system for a $pp$ collision. The timestep is set to be $\delta t=0.005 \ \rm fm/c$. With such a timestep, it is found that the energy drift is less than 1\% for almost all events.

After the initialization, the system begins to evolve. The evolution of system yields color neutral clusters of different sizes due to the interactions among quarks. A cluster is  regarded as well separated from others if the distance between this one and any other quarks outside this cluster is larger than 2 fm. Once all quarks gather into clusters and the numbers and constituents of the clusters don't change over 10 fm/c, the evolution process is considered to be finished. This treatment of hadronization is different from that in Refs. \cite{mol4,mol5}, where a quark-antiquark pair (or three quarks or antiquarks) is regarded as a hadron when the interaction between the pair (or three quarks or antiquarks) and the rest is weaker than a low bound, and the formed hadron is removed from the system.

\begin{figure}[tbph]
\includegraphics[width=0.4\textwidth]{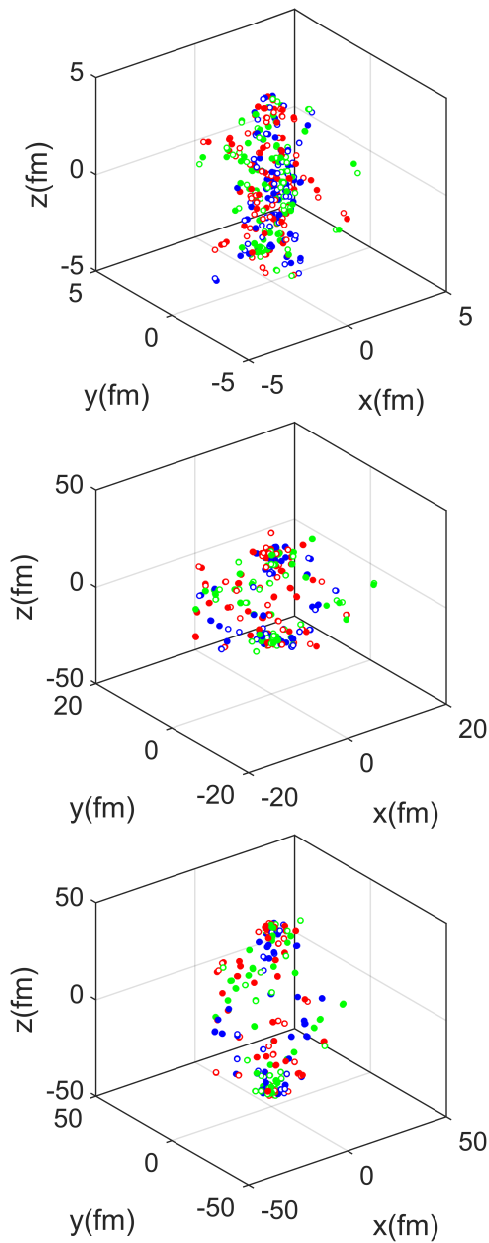}
\caption{(Color online) Evolution of a system with 382 quarks and antiquarks input from PYTHIA at three different points in time: $t=0$ (top panel), $t$=20 fm/c (middle panel) and  $t$=40 fm/c (bottom panel).  The solid and open circles represent quarks and antiquarks respectively, whose colors correspond to the the ones of circles.}
\label{fig_evo}
\end{figure}

A small clusters containing one quark and one antiquark could be mapped as a meson, while  a cluster composed of three quarks (or antiquarks) could be regarded as one baryon (or antibaryon). Occasionally larger clusters containing more than three quarks may also form. These clusters can be regarded as the multi-quark states, such as tetraquark and pentaquark. Because  the relative motions among quarks in the multiquark clusters are usually small (all with speed close to 1), if they happen to be close to each other, with short range interactions, it may be hard for these quarks to be well separated into smaller clusters. Therefore, the multi-quark states may survive long time during the evolution, and are allowed in our model. This is different from the requirement in Refs. \cite{mol5,pol1}.

Fig. \ref{fig_evo} illustrates the quark spatial distributions for an event sample at three different points in time, with the starting time of evolution as $t=0$.

\subsection{Hadron identification}

The mass  $m_h$ of a cluster is determined by its four-momentum,
\begin{eqnarray}
\label{equ5-1}
{m_h}^2 &=&  E_h^2-\vec{p}_h^2 \nonumber \\
&=&\! \! {(\sum_i{\sqrt{m_i^2+\vec{p}_i^2}+\frac{1}{2}\sum_{i \neq j} V_{ij}})}^2\!-{(\sum_i\vec{p}_i)}^2 \; ,
\end{eqnarray}
where the three momentum $\vec{p}_h$ is the sum of the momenta of its constituent quarks, and energy $E_h$ is the sum of the total energy, with $i$ and $j$ the numbers of quarks in the cluster.

Eq. (\ref{equ5-1}) leads to the continuous invariant mass of a cluster, whereas the mass  of the corresponding hadron is a discrete one, which is similar to the scenario in the quark coalescence of AMPT. In identifying a cluster as a hadron, the conservation of three-momentum is chosen. In other words, the three-momentum of one identified hadron is taken as that of the cluster, while its mass is chosen according to its identity. This process will cause a little change of energy for this hadron and for the system.

The criterion for hadron identification is similar to that in PACIAE and AMPT. Formed mesons with the same flavor composition may be of pseudoscaler meson or vector meson. If the invariant mass of a cluster with a quark and an antiquark is nearer to that of a pseudoscaler meson, it is recognized as a pseudoscaler meson, otherwise it is a vector meson. The same principle is used in  the identification of octet and decuplet baryons. As for the formation of flavor-diagonal mesons, the formation probabilities are considered, which can be determined as follows.

For flavor-diagonal clusters containing $u \overline{u}$, $d \overline{d}$, $s \overline{s}$,
the mixing angles are taken to be the same ones used in PYTHIA \cite{pyth1},  which give the wavefunctions of pseudoscalar mesons

\begin{eqnarray}
\label{equ5-2}
\left \{
\begin{aligned}
\pi^0 &= \frac{1}{\sqrt{2}}(u\overline{u}-d\overline{d})\ , \\
\eta &= \frac{1}{2}(u\overline{u}+d\overline{d})-\frac{1}{\sqrt{2}}s\overline{s}\ , \\
\eta ' &= \frac{1}{2}(u\overline{u}+d\overline{d})+\frac{1}{\sqrt{2}}s\overline{s}\ ,
\end{aligned}
\right.
\end{eqnarray}

and vector mesons

\begin{eqnarray}
\label{equ5-3}
\left \{
\begin{aligned}
\rho^0 &= \frac{1}{\sqrt{2}}(u\overline{u}-d\overline{d})\ , \\
\omega &= \frac{1}{\sqrt{2}}(u\overline{u}+d\overline{d})\ , \\
\phi &= s\overline{s} \ .
\end{aligned}
\right.
\end{eqnarray}

Then we rewrite them to get the wavefunctions of $u \overline{u}$, $d \overline{d}$, $s \overline{s}$ represented by pseudoscalar mesons,

\begin{eqnarray}
\label{equ5-4}
\left \{
\begin{aligned}
u \overline{u} &= \frac{1}{2}(\eta+\eta ')+\frac{1}{\sqrt{2}}\pi ^0\ , \\
d \overline{d} &= \frac{1}{2}(\eta+\eta ')-\frac{1}{\sqrt{2}}\pi ^0\ , \\
s \overline{s} &= \frac{1}{\sqrt{2}}(\eta ' -\eta) \ ,  
\end{aligned}
\right.
\end{eqnarray}

and by vector mesons

\begin{eqnarray}
\label{equ5-5}
\left \{
\begin{aligned}
u \overline{u} &= \frac{1}{\sqrt{2}}(\omega+\rho^0) \ ,\\
d \overline{d} &= \frac{1}{\sqrt{2}}(\omega-\rho^0) \ ,\\
s \overline{s} &= \phi \ .  
\end{aligned}
\right.
\end{eqnarray}

We make the rule that, if the invariant mass of a flavor diagonal cluster (i.e. $u \overline{u}$, $d \overline{d}$, or $s \overline{s}$) is larger than $0.5 \ \rm GeV/c^2$, it is identified as a vector meson, and its species  is determined by the probabilities according to Eq. (\ref{equ5-5}). Otherwise, it is a pseudoscalar meson, and its species  is determined by the probabilities according to Eq. (\ref{equ5-4}).

Since the spins of quarks are not considered in the evolution, the differentiation between clusters  composed of quarks with same flavors is determined by their masses only. The exotic hadrons consisting of more than three quarks are not considered in the identification process.

Once a hadron is identified, an id code could be given to it, just as the input procedure for initial state mentioned above. After the identification process, the id number, four-momentum, and mass of a hadron are known. Next these identified hadrons can be again put in PYTHIA to decay to get stable hadrons (while the ``ProcessLevel'' and ``PartonLevel'' in PYTHIA are switched off, and only ``HadronLevel'' is used). Therefore, finally distributions of hadrons produced after decay are influenced not only by the dynamics during evolution process, but also by the decay which is determined by  hadron identification (hence the mass of one cluster to be identified). For now, our main focus is on the influence of the dynamics itself, so the decay process is not considered and only hadrons formed directly from the hadronization are taken into consideration in our study.

\section{Numerical results from the model}

The main tunable parameters in our model are interaction coefficients $a$ and $b$  in Eq. (\ref{equ1-1}) which affect the evolution process and maximum formation time $t_{f,\rm max}$ and time $t_{f}^{'}$ for free flow in Eq. (\ref{equ4-4}) which determine the configuration of the initial state. By setting different parameters, the  distributions and correlations of final  state hadrons could vary.
The influence of parameters is shown in the following numerical results from our model. About 150 thousand non-single diffractive (NSD) events of $pp$  collisions are generated for each kind of combination of parameters to study their influence on the distributions and correlations.

\subsection{Baryon to meson ratios}

Define $N_B$ as the sum of multiplicities of newly produced baryons and antibaryons, and $N_M$ as the multiplicity of mesons. Baryon to meson ratio $R=N_B/N_M$ is studied as a function of the total multiplicity in the whole phase space. The high energy hadrons (with energy larger than 50 GeV) are included in the calculation for comparison.

\begin{figure}[tbph]
\includegraphics[width=0.46\textwidth]{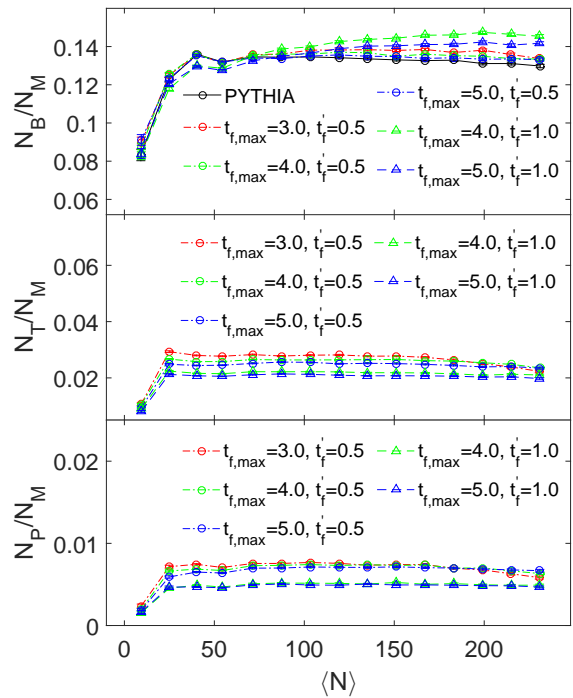}
\caption{(Color online) Multiplicity dependence of baryon to meson (top panel), tetraquark to meson (middle panel)  and pentaquark to meson (bottom panel) ratios  in NSD $pp$ collisions at $\sqrt{s}=7 \rm \ TeV$  for different combinations of maximum formation time $t_{f,\rm max}$ and time $t_{f}^{'}$ for free flow, with fixed $a=0.4 \ \rm GeV/fm$, $b=0.2 \ \rm GeV \cdot fm$. The abscissa values represent the  average total multiplicity of events within each interval. The units of time (fm/c) are omitted in the legends.}
\label{figp1}
\end{figure}

\begin{figure}[tbph]
\includegraphics[width=0.46\textwidth]{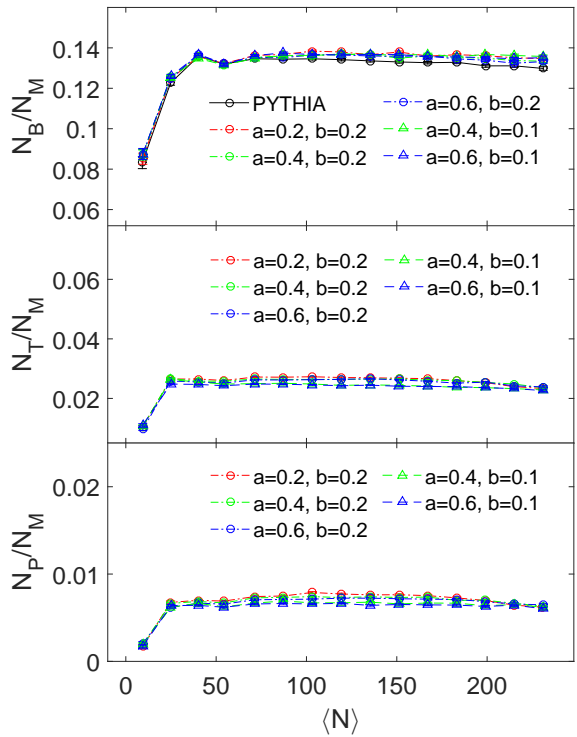}
\caption{(Color online) Multiplicity dependence of baryon to meson (top panel), tetraquark to meson ratio (middle panel)  and pentaquark to meson (bottom panel) ratios    in NSD $pp$ collisions at $\sqrt{s}=7 \rm \ TeV$ for different combinations of interaction coefficients $a$ and $b$, with fixed $t_{f,\rm max}=4.0 \  \rm fm/c$, $t_{f}^{'}=0.5 \ \rm fm/c$.  The units of $a$ and $b$ (GeV/fm and $\rm GeV \cdot fm$ respectively) are omitted in the legends. }
\label{figp2}
\end{figure}

Fig. \ref{figp1} and Fig. \ref{figp2}  present the dependence of particle ratios for different combinations of parameters. The horizontal axes are the total multiplicity in the whole phase space. The top panels in Fig. \ref{figp1} and Fig. \ref{figp2} show the the values of $R$ for different multiplicity event intervals.  The initial baryon number (2 for $pp$ collisions) is excluded in the calculation. The multiquark states are taken into consideration in the calculation. A tetraquark is regarded as a molecular state of two mesons, while a pentaquark is regarded as a molecular state of one meson and one baryon (or antibaryon).  The middle panels present the multiplicity dependence of tetraquark to meson ratios, while the bottom panels pentaquark to meson ratios.

As shown in Fig. \ref{figp1}, the ratios of baryon to meson, tetraquark to meson and pentaquark to meson vary as functions of the total multiplicities for different combinations of  $t_{f,\rm max}$ and $t_f^{'}$, while $a$ and $b$  are fixed at $a=0.4 \ \rm GeV/fm$, $b=0.2 \ \rm GeV \cdot fm$. From the top panel in this figure, one can see that, for each distribution, $R$ rises sharply with the increase of multiplicity  for events with low multiplicity, and then changes much slowly with the increase of multiplicity. For different combinations of $t_{f,\rm max}$ and $t_f^{'}$, the differences between the distributions are more obvious for events with high multiplicities ($\langle N \rangle \gtrsim 100$). The change of $t_f^{'}$ has larger effect than $t_{f,\rm max}$ does. The increase of $t_{f,\rm max}$ reduces the ratios for events with high multiplicity, but the ratios change little for events with low multiplicity ($\langle N \rangle \lesssim 80$). The increase of $t_f^{'}$, by contrast, increases the ratios for high multiplicity events, while decreases them for events with low multiplicity. In the middle panel, for each distribution the tetraquark to meson ratio shows a sharp rise for low multiplicity events ($\langle N \rangle \lesssim 25$), then slowly drops with the increase of multiplicity. The ratios decrease with the increase of $t_{f,\rm max}$ or $t_f^{'}$ for most multiplicity intervals. As for the pentaquark to meson ratios shown in the bottom panel, the ratios are much smaller than  the former two and change little for most events with different $t_{f,\rm max}$, and show slight decrease with the increase of $t_f^{'}$.

Fig. \ref{figp2} presents the same ratios as in Fig. \ref{figp1} for different combinations of interaction coefficients $a$ and $b$, while $t_{f,\rm max}$ and $t_f^{'}$ are fixed, $t_{f,\rm max}=4.0 \  \rm fm/c$, $t_{f}^{'}=0.5 \ \rm fm/c$. As is depicted in the top panel, $R$ shows a tendency similar to that in Fig. \ref{figp1} for each distribution, and the differences between distributions  with different combinations of $a$ or $b$ are more obvious for events with high multiplicities, which resembles to the above case. For high multiplicity events ($\langle N \rangle \gtrsim 170$ ), the increase of $a$ or $b$ slightly decreases the ratios, but the effect of $b$ is a little more obvious. For events with lower multiplicity, the change of $a$ or $b$ has little effect on the change of ratios.  As is shown in the middle panel, the tetraquark/meson ratio also shows a sharp rise for low multiplicity events ($\langle N \rangle \lesssim 25$), then slowly drops with the increase of multiplicity for each distribution. The change of $a$ doesn't have evident effect, while the increase of $b$ slightly increases the ratios for most high multiplicity intervals ($\langle N \rangle \gtrsim 80$). For the pentaquark/meson ratio shown in the bottom panel, it shows a sharp rise for low multiplicity events ($\langle N \rangle \lesssim 25$), then slowly drops with the increase of multiplicity for each distribution. The change of $a$ doesn't have evident effect. The increase of $b$ increases the ratios slightly for most high multiplicity intervals ($\langle N \rangle \gtrsim 80$).

\subsection{Transverse momentum distributions}
\label{subsec1}

\begin{figure}[tbph]
\includegraphics[width=0.46\textwidth]{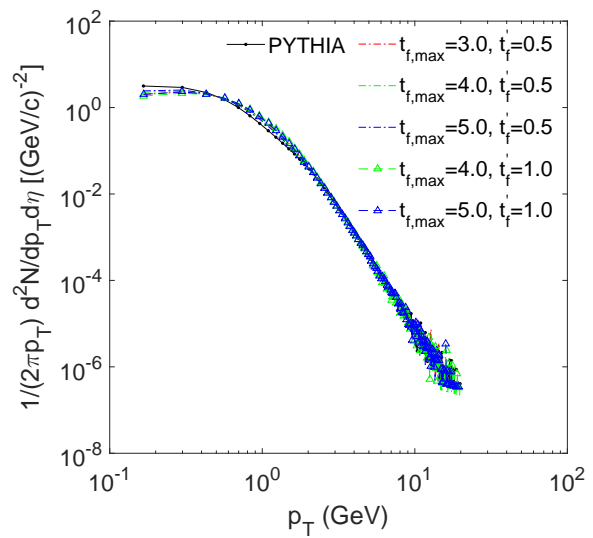}
\caption{(Color online) Transverse momentum distribution in $|\eta|<0.5$ region   in NSD $pp$ collisions at $\sqrt{s}=7 \rm \ TeV$ for different combinations of maximum formation time $t_{f,\rm max}$ and time $t_{f}^{'}$ for free flow, with fixed $a=0.4 \ \rm GeV/fm$, $b=0.2 \ \rm GeV \cdot fm$. The units of time (fm/c) are omitted in the legend.}
\label{fig1}
\end{figure}

\begin{figure}[tbph]
\includegraphics[width=0.46\textwidth]{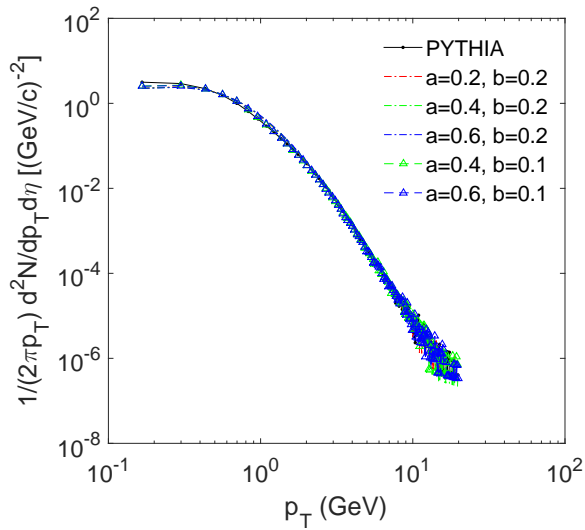}
\caption{(Color online) Transverse momentum distribution in $|\eta|<0.5$ region  in NSD $pp$ collisions at $\sqrt{s}=7 \rm \ TeV$ for different combinations of interaction coefficients $a$ and $b$, with fixed $t_{f,\rm max}=4.0 \ \rm fm/c$, $t_{f}^{'}=0.5 \ \rm fm/c$.  The units of $a$ and $b$ (GeV/fm and $\rm GeV \cdot fm$ respectively) are omitted in the legend. }
\label{fig2}
\end{figure}

Fig. \ref{fig1} and Fig. \ref{fig2} show the dependence of transverse momentum  distributions for different values of parameters in the mid-pseudorapidity region $|\eta|<0.5$.

Fig. \ref{fig1} presents  the dependence of transverse momentum distribution for different combinations of maximum value of formation time $t_{f,\rm max}$ and free flow time $t_{f}^{'}$, while $a$, $b$ are fixed, $a=0.4 \ \rm GeV/fm$, $b=0.2 \ \rm GeV \cdot fm$. One can see that the transverse momentum distributions for hadrons within $p_T\lesssim$ 2.0 GeV/c are evidently affected. This can be  attributed to the expansion in the transverse plane, which makes the interactions have more evident effect on the transverse motions of quarks. As a result, more quarks tend to increase their $p_T$, which leads the decrease of hadron number in the small $p_T$ region and the increase in the large $p_T$ region.  With the increase of $t_{f,\rm max}$, the deviation of distribution from the result of PYTHIA  slightly decreases. Conversely, the deviation increases with the increase of $t_{f}^{'}$. This can be explained as follows. Because the total color of quarks from the same parent hadron is neutral, then forces acting on one quark from quarks decomposed from the same parent hadron could cancel each other out to some degree, which is determined by the values of formation time and free flow time. The increase of $t_{f,\rm max}$ makes more quarks from different parent hadrons separated in space or increases the separating distances, and thus weakens the interactions between quarks from different parent hadrons, while the increase of $t_{f}^{'}$ creates the opposite effect by having more possibility to mix more quarks from different parent hadrons. In other words, the values of  $t_{f,\rm max}$ and $t_{f}^{'}$ influence the effective range of interactions. For larger $t_{f,\rm max}$, the interactions between quarks from different parent quarks are weaker, which makes the transverse motions of quarks tend to be less affected by other quarks during the expansion process. A larger $t_{f}^{'}$ enhances the range of interactions, but also reduces the short-range Coulombic interactions. However, from the results, one can conclude that the former one is stronger, which means a larger $t_{f}^{'}$ will enhance the interactions and tend to increase the $p_T$ of quarks.

The influence of different values of $a$ and $b$ is shown in Fig. \ref{fig2}. The difference between the distributions from PYTHIA and our model increases with the increase of $b$, while changes little with the variation of $a$. Clearly $b$ has larger effect on the distribution, since it determines the strength of short range interaction in Eq. (\ref{equ1-1}), which is much larger than the linear one when quarks are close to  each other. This indicates that the deviation is mainly formed in the early stage of the evolution process of a system,  when the quarks in the system are compact, and the Coulombic potential, as short range interaction, is of great influence.

\subsection{Pseudorapidity distributions}

\begin{figure}[tbph]
\includegraphics[width=0.48\textwidth]{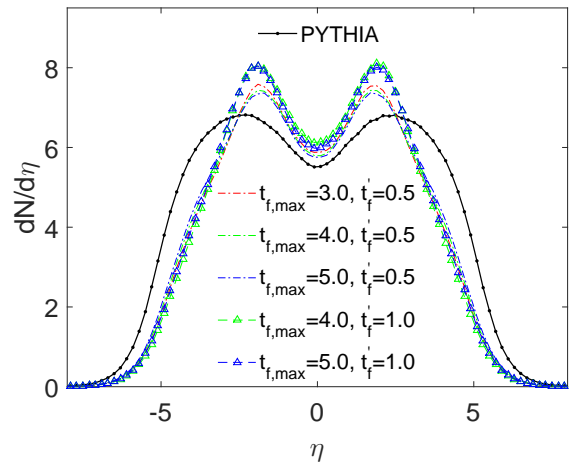}
\caption{(Color online) Pseudorapidity distribution in NSD $pp$ collisions at $\sqrt{s}=7 \rm \ TeV$ for different combinations of maximum formation time $t_{f,\rm max}$ and time $t_{f}^{'}$ for free flow, with fixed $a=0.4 \ \rm GeV/fm$, $b=0.2 \ \rm GeV \cdot fm$. The units of time fm/c are omitted in the legend.}
\label{fig3}
\end{figure}

\begin{figure}[tbph]
\includegraphics[width=0.48\textwidth]{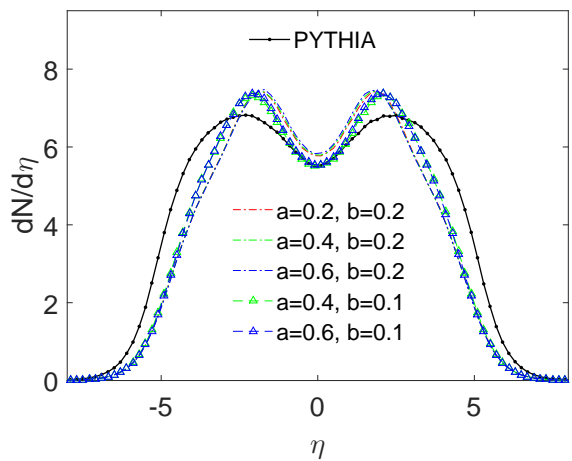}
\caption{(Color online) Pseudorapidity distribution in NSD $pp$ collisions at $\sqrt{s}=7 \rm \ TeV$ for different combinations of interaction coefficients $a$ and $b$, with fixed $t_{f,\rm max}=4.0 \ \rm fm/c$, $t_{f}^{'}=0.5 \ \rm fm/c$.  The units of $a$ and $b$ (GeV/fm and $\rm GeV \cdot fm$ respectively) are omitted in the legend. }
\label{fig4}
\end{figure}

Fig. \ref{fig3} and Fig. \ref{fig4} show the pseudorapidity distributions for different values of parameters. There is no restriction on the  selection of transverse momentum.

As is shown in Fig. \ref{fig3}, with fixed $a$ and $b$, the pseudorapidty distributions are evidently  deviated from the one from PYTHIA. The distributions in pseudorapidity region $|\eta|\gtrsim 2.0$ decrease remarkably, while distributions in $|\eta|\lesssim 2.0$ show obvious increase. Such a change can be accounted for by the increase of transverse momentum as a result of the interactions. Quarks with large pseudorapidities are often the ones with small transverse momenta or large longitudinal momenta, therefore they are very sensitive to the forces acting on them. From Fig. \ref{fig_evo}, one can see the quarks with large speeds (usually with large momenta) are often crowded in the forward/backward direction, and thus they are strongly affected by the interactions. The decrease of the maximum value of formation time $t_{f,\rm max}$ or increase of free flow time $t_{f}^{'}$ leads to a larger deviation of the distribution from  that from PYTHIA. The explanation for such change is the same as the one mentioned in Subsection ~{\ref{subsec1}} for the transverse momentum distributions. Additionally, the change of $t_{f}^{'}$  has larger effect on the pseudorapidity distribution than  $t_{f,\rm max}$ does.

From Fig. \ref{fig4}, one can see, with fixed $t_{f,\rm max}$  and $t_{f}^{'}$, for different combinations of $a$ and $b$, the distributions deviate from that from PYTHIA. Similar to the former scenario, the distributions in pseudorapidity region $|\eta|\gtrsim 2.0$ decrease remarkably, while in $|\eta|\lesssim 2.0$ show obvious increase. The increase of $b$ could obviously enlarge the deviation, while the increase of $a$ only has a much smaller effect on the increase of the deviation, which indicates that the deviations of distributions  from that of PYTHIA mainly stem from the initial stage of the evolution when the quarks are more compact.

\begin{figure}[tbph]
\includegraphics[width=0.48\textwidth]{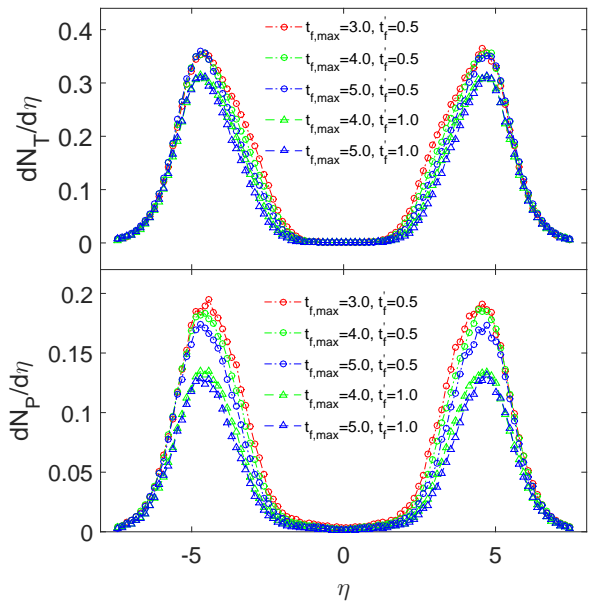}
\caption{(Color online) Pseudorapidity distribution of tetraquarks (top panel) and pentaquarks (bottom panel)  in NSD $pp$ collisions at $\sqrt{s}=7 \rm \ TeV$ for different combinations of maximum formation time $t_{f,\rm max}$ and time $t_{f}^{'}$ for free flow, with fixed $a=0.4 \ \rm GeV/fm$, $b=0.2 \ \rm GeV \cdot fm$. The units of time fm/c are omitted in the legends.}
\label{figp3}
\end{figure}

\begin{figure}[tbph]
\includegraphics[width=0.48\textwidth]{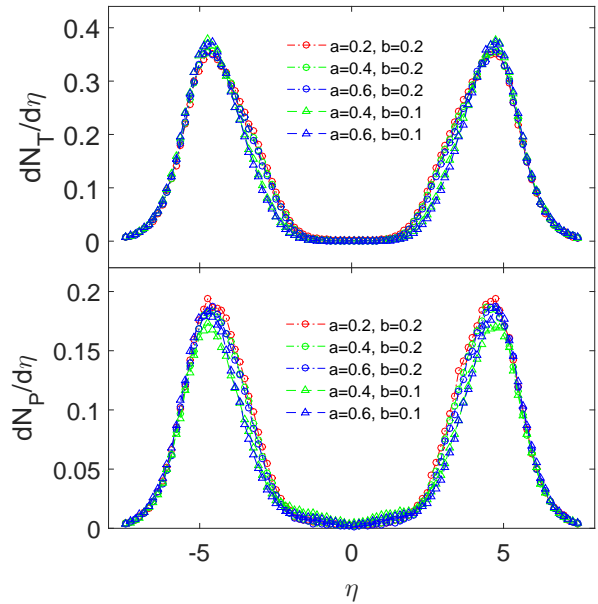}
\caption{(Color online) Pseudorapidity distribution of tetraquarks  (top panel) and pentaquarks  (bottom panel)  in NSD $pp$ collisions at $\sqrt{s}=7 \rm \ TeV$ for different combinations of interaction coefficients $a$ and $b$, with fixed maximum value of formation time $t_{f,\rm max}=4.0 \ \rm fm/c$, and free flow time $t_{f}^{'}=0.5 \ \rm fm/c$.  The units of $a$ and $b$ (GeV/fm and $\rm GeV \cdot fm$ respectively) are omitted in the legends. }
\label{figp4}
\end{figure}

Fig. \ref{figp3} and Fig. \ref{figp4}  show the pseudorapidity distributions of tetraquark and pentaquark respectively for different combinations of parameters. One can see from those figures that the multiquark states are mainly distributed in the very forward/backward pseudorapidity region. This can be understood, since quarks in the very forward/backward pseudorapidity region are often the ones with large momenta, which makes the quarks more easily stick together if they happen to be close to each other from the beginning of the evolution, and thus  have a larger possibility to form multi-quark states.

\subsection{Forward-backward multiplicity correlations}

Correlations and fluctuations are important tools to study the mechanism of particle production in high energy collisions. The correlations and fluctuations of produced particles may change significantly in a phase transition because of the change of degree of freedom. Forward-backward  multiplicity correlations have been studied widely to analyze the mechanism of particle production  \citep{fbcorr1,fbcorr2}  and played an important role during the development of mechanism of multi-parton interactions in PYTHIA  model \citep{pythmpi}.  The forward-backward  multiplicity correlations are usually defined as the correlation coefficients of forward and backward multiplicities

\begin{eqnarray}
\label{equ6-1}
b_{corr}&=&\frac{\langle n_F n_B \rangle -\langle n_F\rangle \langle n_B \rangle }{ \sqrt{(\langle n_F^2 \rangle - \langle n_F \rangle ^2)(\langle n_B^2 \rangle - \langle n_B \rangle ^2)}} \ ,
\end{eqnarray}
where $n_F$ and $n_B$  are the numbers of particles in the forward and backward pseoduorapidity intervals which are symmetrically located in pseudorapidity. For symmetric distributions (e.g. those from $pp$ collisions),  Eq. (\ref{equ6-1}) could be rewritten as
\begin{eqnarray}
\label{equ6-2}
b_{corr}=\frac{\langle n_F n_B \rangle -\langle n_F\rangle ^2 }{ \langle n_F^2 \rangle - \langle n_F \rangle ^2}\ .
\end{eqnarray}

\begin{figure}[tbph]
\includegraphics[width=0.48\textwidth]{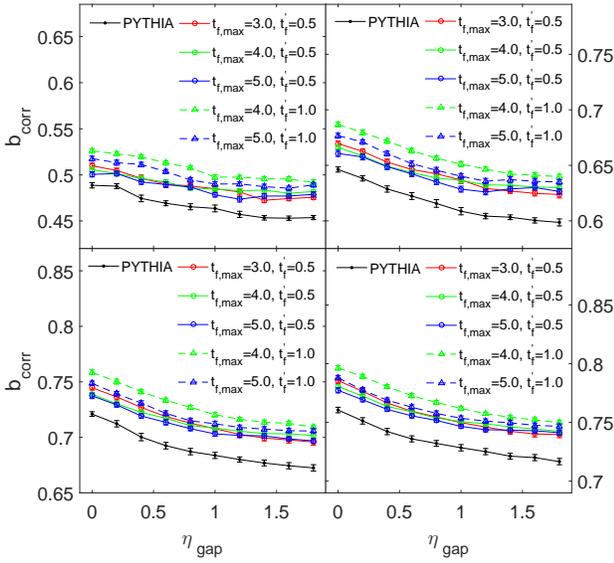}
\caption{(Color online) Forward-backward correlations in NSD $pp$ collisions at $\sqrt{s}=7 \rm \ TeV$ for different combinations of maximum formation time $t_{f,\rm max}$ and time $t_{f}^{'}$ for free flow, with fixed $a=0.4 \ \rm GeV/fm$, $b=0.2 \ \rm GeV \cdot fm$. $\delta \eta$  is the size of intervals selected symmetrically, with    $\eta_{gap}$ as the variable central separation. The units of time fm/c are omitted in the legends.}
\label{fig5}
\end{figure}

\begin{figure}[tbph]
\includegraphics[width=0.48\textwidth]{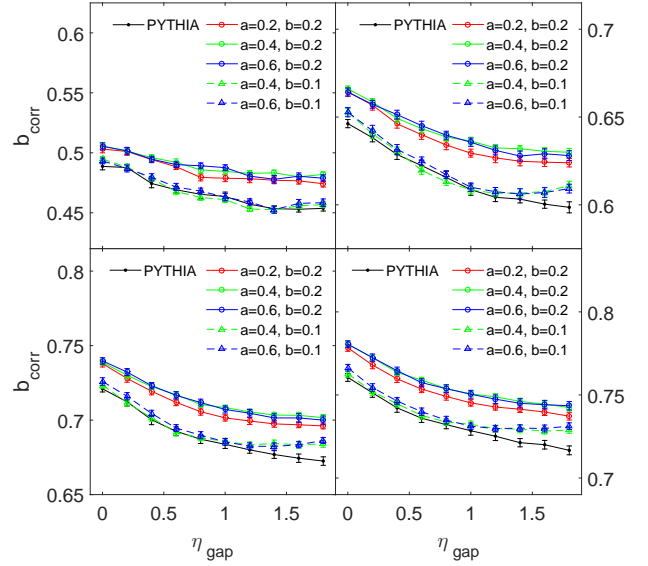}
\caption{(Color online) Forward-backward correlations  in NSD $pp$ collisions at $\sqrt{s}=7 \rm \ TeV$ for different combinations of interaction coefficients $a$ and $b$, with fixed $t_{f,\rm max}^{'}=4.0 \ \rm fm/c$, $t_{f}^{'}=0.5 \ \rm fm/c$. $\delta \eta$  is the size of intervals selected symmetrically, with    $\eta_{gap}$ as the variable central separation.  The units of $a$ and $b$ (GeV/fm and $\rm GeV \cdot fm$) are omitted in the legends. }
\label{fig6}
\end{figure}

In our study, we choose the forward and backward pseudorapidity regions symmetrically with a pseudorapidity gap $\eta_{gap}$ and width $\delta \eta$ for each side. The results of correlation strength from our model are about ten percent larger than those from the original PYTHIA model and are shown in Fig. \ref{fig5} and Fig. \ref{fig6} for different sets of parameters. As depicted in these figures, the  results show a decrease of correlation strength with the increase of pseudorapidty gap, and an increase with the increase  of interval size.

Fig. \ref{fig5} presents the dependence of forward-backward correlations as a function of $\eta_{gap}$ and $\delta \eta$ for different values of maximum formation time $t_{f,\rm max}$ and free flow time $t_{f}^{'}$ with fixed $a=0.4 \ \rm GeV/fm$, $b=0.2 \ \rm GeV \cdot fm$. The change of $t_{f}^{'}$ has larger effect than  $t_{f,\rm max}$ does. The increase of $t_{f,\rm max}$ slightly decreases correlations for different $\delta \eta$ for most pseudorapidity gaps, whereas the increase of free flow time obviously enhances the correlations with different pseudorapidity gaps and widths. This can be explained by considering that the increase of free flow time makes quarks from different parent hadrons closer in space, 
while the increase of $t_{f,\rm max}$ tends to weaken such mixing. Such mixing will enhance the correlations of final hadrons.

The dependence of the correlations on parameters $a$ and $b$ is presented in Fig. \ref{fig6}. The increase of $b$ leads to the increase of the correlations, whereas the change of $a$ has little effect on the correlations, which indicates that the contribution to correlations is mainly from the initial interactions  when the partons are close and the Coulombic term in Eq. (\ref{equ1-1}) plays a major role in the evolution process.

\section{Summary and discussions}

In this paper, a hadronization mechanism based on color interactions among relativistic quarks is discussed and applied to study the hadronization for high energy $pp$ collisions.  The distributions and forward-backward correlation strength from simulations with different parameters are compared with those from PYTHIA. Our study focuses on the effects of the dynamics controlled by parameters on the final distributions and correlations. The initial geometry is determined according to a method inspired from AMPT model, while the interacting potential is of Cornell type. The effects of the color interactions, maximum formation time and free flow time  are discussed. As revealed from the results above, free flow time  and Coulombic interaction have much more evident effects on all the hadron distributions and correlations.

The present work can be improved from several aspects: (1) The initial space positions of hadrons and quarks should stem from some more solid basis; (2) The form of interaction can include more realistic physical effects, such as Debye screening etc.; (3) Relativistic retarding effect can also be considered. All these can be done in later work.

\begin{acknowledgments}

This work was supported in part by National Natural Science Foundation of China (11435004, 11375069), the Ministry of Science and Technology of China (2015CB56901), and the Programme of Introducing Talents of Discipline to Universities (B08033).

\end{acknowledgments}


\begin{thebibliography}{99}
\vspace{3mm}
\bibitem{mod1} B. Andersson, G. Gustafson, G. Ingelman and T. Sjostrand, Phys. Rept. {\bf 97}, 31-145 (1983).

\bibitem{mod2} 
B. Andersson, S. Mohanty, F. Soderberg, aXiv: hep-ph/0212122 .

\bibitem{mod3}  
T. Sjostrand, Nucl. Phys. B {\bf 248}, 469 (1984).

\bibitem{mod4}
B. Andersson, G. Gustafson and B. Soderberg, Z. Phys. C: Particles and Fields {\bf 20}, 317-329 (1983).

\bibitem{mod5} 
S. Ferreres-Sole, T. Sjostrand, Eur. Phys. J. C {\bf 78}, 983 (2018), arXiv:1808.04619 [hep-ph].

\bibitem{pyth1} 
T. Sjostrand, S. Mrenna and P. Z. Skands, JHEP {\bf 05}, 026 (2006), arXiv:hep-ph/0603175v2 .

\bibitem{pyth2} 
T. Sjostrand, S. Ask, J. R. Christiansen et al, Comput. Phys. Commun. {\bf 191}, 159-177 (2015).

\bibitem{herwig1}
A. Kupco, arXiv: hep-ph/9906412 .

\bibitem{herwig2}  
A. Buckley J. Butterworth, S. Gieseke et al, Phys. Rept.  {\bf 504}, 145-233 (2011), arXiv:1101.2599[hep-ph].


\bibitem{coale1}
R. C. Hwa and C. B. Yang, Phys. Rev. C {\bf 67}, 034902 (2003).

\bibitem{coale2}
V. Greco, C. M. Ko and P. Levai, Phys. Rev. Lett. {\bf 90}, 202302 (2003).

\bibitem{coale3}
R. J. Fries, B. Mueller and C. Nonaka el al, Phys. Rev. Lett. {\bf 90}, 202303 (2003).

\bibitem{coale4}
K. C. Han, R. J. Fries, and C. M. Ko, Phys. Rev. C  {\bf 93}, 045207 (2016).

\bibitem{ampt1}
Zi-Wei Lin, Che Ming Ko, Bao-An Li et al, Phys. Rev. C {\bf 72}, 064901 (2005).


\bibitem{paciae1}
Ben-Hao Sa, Dai-Mei Zhou, Yu-Liang Yan et al., Comput. Phys. Commun. {\bf 183}, 333-346 (2012).

\bibitem{paciae2}
Ben-Hao Sa, Dai-Mei Zhou, Yu-Liang Yan et al., Comput. Phys. Commun. {\bf 184}, 1476-1479 (2013).

\bibitem{paciae3}
Dai-Mei Zhou,  Yu-Liang Yan, Xing-Long Li et al., Comput. Phys. Commun. {\bf 193}, 89-94 (2015).

\bibitem{dy1} Michael P. Allen, NIC Series {\bf 23}, 7 (2004).

\bibitem{dy2} Michael P. Allen, AIP Conference Proceedings {\bf 1091}, 1 (2009).

\bibitem{dy3} 
G. Kresse, J. Hafner, J. Non-Cryst. Solids {\bf 47}, 558-561 (1993).

\bibitem{dy4} K. Albe, K. Nordlund,  R. S. Averback, Phys. Rev. B {\bf 65}, 195124(2002).

\bibitem{dy5} M. N. Offman,  M. Krol, I. Silman et al, J. Biol. Chem. {\bf 285}, 42105-42114 (2010). 

\bibitem{mol1}
P. Hartmann, Z. Donko, P. Levai, G. J. Kalman, Nucl. Phys. A {\bf 774}, 881-884 (2006).

\bibitem{mol2}
Y. Akimura, T. Maruyama, N. Yoshinaga, eta al, Eur. Phys. J. A {\bf 25}, 405-411 (2005).

\bibitem{mol3} R. Marty, J. Aichelin, Phys. Rev. C {\bf 87}, 034912 (2013).

\bibitem{mol4}
S. Scherer, M. Hofmann, M. Bleicher et al., New J. Phys. {\bf 3}, 8 (2001).

\bibitem{mol5}
M. Hofmann,  M.  Bleicher, S. Scherer et al, Phys. Lett. B {\bf 478}, 161-171 (2000).

\bibitem{mol6}
S. Terranova, D. M. Zhou, A. Bonasera, Eur. Phys. J. A {\bf 26}, 333-337 (2005).

\bibitem{mol7} 
A. Bonasera, arXiv:nucl-th/9905025.

\bibitem{pol1} Z. G. Tan, C. B. Yang, Int. J. Mod. Phys. E {\bf 24}, 1550044 (2015).

\bibitem{njl1}
Y. Nambu and G. Jona-Lasinio, Phys. Rev. {\bf 122}, 345 (1961).

\bibitem{njl2}
C. M. Ko, T. Song, F. Li et al, Nucl. Phys. A {\bf 928}, 234-246 (2014), arXiv:1211.5511 [nucl-th].

\bibitem{njl3}
F. Li and C. M. Ko, Phys. Rev. C {\bf 95}, 055203 (2017), arXiv:1606.05012 [nucl-th].

\bibitem{njl4}
F. Li and C. M. Ko, Phys. Rev. C {\bf 93}, 035205 (2016), arXiv:1601.00026 [nucl-th].


\bibitem{cornell1} E. Eichten, K. Gottfried, T. Kinoshita, K. D. Lane, and T. M. Yan, Phys. Rev. D {\bf 17}, 3090 (1978).

\bibitem{cornell2} X. W. Liu, H. W. Ke, X. Liu et al, Phys. Rev. D {\bf 93}, 074013 (2016),  arXiv:1602.00226 [hep-ph].

\bibitem{pot1}
Hee Sok Chung, Jungil Lee, Daekyoung Kang, arXiv:0803.3116[hep-ph].

\bibitem{pot2}
L. P. Fulcher, Z. Chen, K. C. Yeong, Phys. Rev. D {\bf 47}, 4122-4132 (1993).

\bibitem{pot3} 
L. P. Fulcher, Phys.Rev. D {\bf 60}, 074006 (1999).

\bibitem{pot4} P. Gonzalez, J. Phys. G: Nucl. Part. Phys. {\bf 41}, 095001 (2014).


\bibitem{rel1} 
A. Aliano, L. Rondoni, and G. P. Morriss, Eur. Phys. J. B {\bf 50}, 361-365 (2006).

\bibitem{rel2} 
M. Ghodrat, A. Montakhab, Comput. Phys. Commun. {\bf 182}, 1909-1913 (2011).

\bibitem{rel3} J. Dunkel, P. Hanggi, Physica A {\bf 374}, 559 (2007).

\bibitem{pdg}
C. Patrignani et al. (Particle Data Group), Chin. Phys. C {\bf 40}, 100001 (2016).

\bibitem{resc1}
T. J. Humanic, Phys. Rev. C {\bf 79}, 044902 (2009).

\bibitem{resc2}
D. Truesdale, T. J. Humanic, J. Phys. G {\bf 39}, 015011 (2012).

\bibitem{fbcorr1}
E. Dominguez-Rosas, E. C. Flores, EPJ Web Conf. {\bf 172}, 05008 (2018).

\bibitem{fbcorr2}
J. Adam et al. (ALICE Collaboration)., JHEP {\bf  1505}, 097 (2015).

\bibitem{pythmpi}
T. Sjostrand, Adv. Ser. Direct. High Energy Phys. {\bf 29}, 191-225 (2018), arXiv:1706.02166 [hep-ph].


\end{thebibliography}
\end{document}